# High Entropy Alloy Property Predictions Using a Transformer-based Language Model

S.Kamnis[1,2*], K.Delibasis[1]

[1]Department of Computer Science and Biomedical Informatics, University of Thessaly, 35100 Lamia, Greece

[2]Castolin Eutectic-Monitor Coatings Ltd., Newcastle upon Tyne NE29 8SE, UK

*Corresponding Author

**Abstract**

This study introduces a language transformer-based machine learning model to predict key mechanical properties of high-entropy alloys (HEAs), addressing the challenges due to their complex, multi-principal element compositions and limited experimental data. By pre-training the transformer on extensive synthetic materials data and fine-tuning it with specific HEA datasets, the model effectively captures intricate elemental interactions through self-attention mechanisms. This approach mitigates data scarcity issues via transfer learning, enhancing predictive accuracy for properties like elongation (%) and ultimate tensile strength (UTS) compared to traditional regression models such as Random Forests and Gaussian Processes. The model's interpretability is enhanced by visualizing attention weights, revealing significant elemental relationships that align with known metallurgical principles. This work demonstrates the potential of transformer models to accelerate materials discovery and optimization, enabling accurate property predictions, thereby advancing the field of materials informatics.

**Introduction**

High-entropy alloys (HEAs) are an innovative class of materials distinguished by their multi-principal element compositions, typically consisting of five or more elements in near-equiatomic ratios. Unlike traditional alloys, which are based on one or two primary elements, HEAs utilize high configurational entropy to stabilize simple solid-solution phases instead of intermetallic compounds. This unique compositional approach results in exceptional properties, including high strength, excellent ductility, superior wear resistance, and remarkable thermal stability. These attributes make HEAs promising candidates for advanced applications in the aerospace, automotive, and energy sectors. Designing HEAs involves navigating an extensive compositional space due to the vast number of possible element combinations and concentrations [1][2][3][4][5][6][7][8][9][10].

Traditional design methods rely on phase diagrams and atomistic simulations to predict phase stability and material properties. Phase diagrams are essential for understanding equilibrium phases and transformations; however, they become less reliable when extrapolating beyond known Gibbs energies. In multi-component systems, interpolating Gibbs energies becomes increasingly challenging as the number of elements increases, complicating the prediction of phase formations in unexplored compositional regions. Atomistic simulations, such as those based on Density Functional Theory (DFT), provide detailed insights into the electronic structure and thermodynamic properties of materials. DFT can predict phase stability, mechanical properties, and electronic behaviour from first principles. However, the computational cost of DFT grows significantly with system size and complexity. For HEAs, which involve multiple principal elements and complex crystal structures, constructing accurate DFT models requires large supercells to capture the inherent disorder and configurational entropy, making such calculations computationally prohibitive [11][12][13][14][15]

To address these challenges, machine learning (ML) algorithms have been employed to predict the properties and phase formations of HEAs, thereby guiding experimental efforts more efficiently. Conventional ML models, including artificial neural networks (ANNs) [16][17][18][19][20][21][22][23] , support vector machines (SVMs) [24][25][26], Gaussian Process (GP) [27][28][29][30][31][32], k-nearest neighbours (KNN) [33][34] and random forests (RFs) [35][36], have been used to correlate elemental features and processing parameters with material properties. Advanced algorithms, such as deep learning (DL) models—including deep neural networks (DNNs) and convolutional neural networks (CNNs) [37][38] [39] have also been applied to capture the complex nonlinear relationships present in HEA systems.

Despite these advancements, ML applications in HEA design encounter significant obstacles. A primary challenge is the reliance on large, high-quality datasets, which are often limited due to experimental constraints. Insufficient data prevents the training of robust ML models, often leading to overfitting and poor generalization on unseen compositions. Additionally, traditional ML models may struggle to capture the high-dimensional feature spaces and long-range elemental interactions characteristic of HEAs. The "black-box" nature of complex ML models also poses interpretability issues, making it difficult to discern the underlying factors influencing predictions. To overcome these limitations, we propose a new approach that involves pre-training a transformer-based model on extensive materials data and fine-tuning it with experimental data specific to HEAs. Transformers, originally developed for natural language processing tasks [40][41], utilize self-attention mechanisms to model complex relationships within sequences, effectively capturing long-range dependencies and interactions. By pre-training on large-scale materials datasets, the transformer model learns generalized representations of elemental properties and interactions. Fine-tuning with experimental HEA data allows the model to adapt these representations to the specific complexities of HEAs, enhancing its predictive capabilities even with limited data [42].

Our approach offers several advantages over traditional methods and addresses some of the pressing challenges in the field. By leveraging pre-trained models, we mitigate data scarcity issues through the transfer of knowledge from larger, accurately calculated datasets that do not rely solely on experimental data. This transfer significantly improves model performance on small HEA datasets, enhancing data efficiency. The transformer's ability to capture complex, high-dimensional relationships enhances its generalization to new, unseen alloy compositions, facilitating the discovery of novel HEAs with desired properties. Moreover, unlike DFT simulations, our method does not require extensive computational resources once the model is pre-trained, making it more practical for screening large compositional spaces. Additionally, the self-attention mechanism, inherent in transformers, provides valuable insights into feature importance and elemental interactions, addressing the interpretability challenges often associated with traditional deep learning models. Furthermore, fine-tuning allows the model to be easily adapted to different HEA systems or target properties without the need to retrain from scratch, enhancing its adaptability and making it a versatile tool for materials design.

In this work, we compare our proposed transformer-based approach with traditional regression models such as Random Forests (RF), Gaussian Processes (GP), and Gradient Boosting in their ability to predict two macroscopic mechanical properties: elongation at break and ultimate tensile strength (UTS). While regression models may perform variably depending on the task and dataset size, the proposed model consistently achieves higher performance across different tasks, offering universal applicability by integrating the strengths of various models into a single framework. We assess the impact of pre-training dataset size on model performance, confirming that larger datasets enhance the model's capability to capture complex elemental interactions. Additionally, we employ interpretability techniques like attention weight visualization to elucidate the model's decision-making process. This work builds on our previous short communication publication [43] and serves as a proof of concept for applying transformer-based models to HEA design. We recognize that further enhancements to the pre-training datasets, particularly by incorporating more and diverse thermodynamic properties, could significantly enrich the model's understanding of material behaviours. Additionally, experimenting with different transformer architectures and large language models (LLMs) may prove pivotal in achieving even greater predictive performance. These efforts could refine the model's ability to capture intricate elemental interactions and thermodynamic principles, ultimately accelerating the discovery of HEAs with optimized properties.

**Data and Methodology**

A comprehensive comparison has been undertaken for various regression models to predict material properties using two distinct datasets: Ultimate Tensile Strength (UTS) and elongation. The UTS dataset is characterized by its relative simplicity, reduced experimental uncertainties and larger dataset making it relatively straightforward for predictive modelling. In contrast, the elongation dataset presents a more formidable challenge due to smaller dataset size, inherent measurement errors and noise, reflecting real-world complexities where data imperfections are commonplace.

A critical aspect of this work is the utilization of raw, experimental-driven data without any preprocessing steps such as cleansing and outlier detection. Traditionally, these preprocessing techniques are employed to enhance model accuracy by mitigating the impact of anomalies and noise. However, our deliberate omission of these steps serves a dual purpose: it allows us to assess how different models inherently handle noisy and imperfect data, and it provides insights into their robustness in less controlled environments.

The selection of regression models was based on popularity as state-of-the-art in the field of HEA material informatics while encompassing algorithms known for their efficacy with varying dataset sizes. Some models, like Random Forests and Gradient Boosting Regressors, are renowned for their superior performance with large, complex datasets due to their ensemble learning capabilities. Conversely, models such as Gaussian Process, Support Vector Regression and K-Nearest Neighbours are often preferred for smaller datasets where overfitting is a concern.

**Dataset Analysis-Supervised Learning**

The elongation and UTS data are for as-cast HEAs tested under both compression and tension and at Room temperature. The dataset is available at: https://github.com/SPS-Coatings/Language-Model-for-HEA

**Elongation Fine Tuning Dataset**

The bar chart in Figure 1.d shows the total amount of each element present in the dataset, with Ni, Fe, Co, and Cr being dominant. This overrepresentation may introduce bias into the model, as the predictions could skew toward alloys with these elements, reducing the accuracy for compositions with less frequent elements like Si, Sc, or Sn. This imbalance could limit the generalizability of the model to diverse alloy compositions. The histogram (Figure 1.b) for elongation percentages reveals a broad distribution, with peaks around 10% and 40%, showing variability in elongation behaviour across different compositions. The correlation heatmap (Figure 1.a) provides additional insight into feature importance. While properties like modulus mismatch and ionization energy show moderate correlation with elongation, many other features (e.g., number of elements, melting temperature) show weak or negative correlations. These weak correlations suggest that compositional features alone may not be sufficient to predict elongation accurately, as elongation is influenced by further microstructural characteristics. Figure 1.c chart shows the distribution of the number of elements per composition revealing that most alloys contain 5 or 6 elements as expected for a HEA focused task. This skew may lead to biased model performance for alloys with 7+ elements, which are underrepresented. In conclusion, the dataset's small size, compositional imbalance, and non-linear trends in elongation introduce significant challenges for a regression model. The model is likely to struggle with generalizing to less frequent element combinations. For a production ready predictive model for this target property, additional data and are required to improve prediction accuracy. This is out of this work's scope.

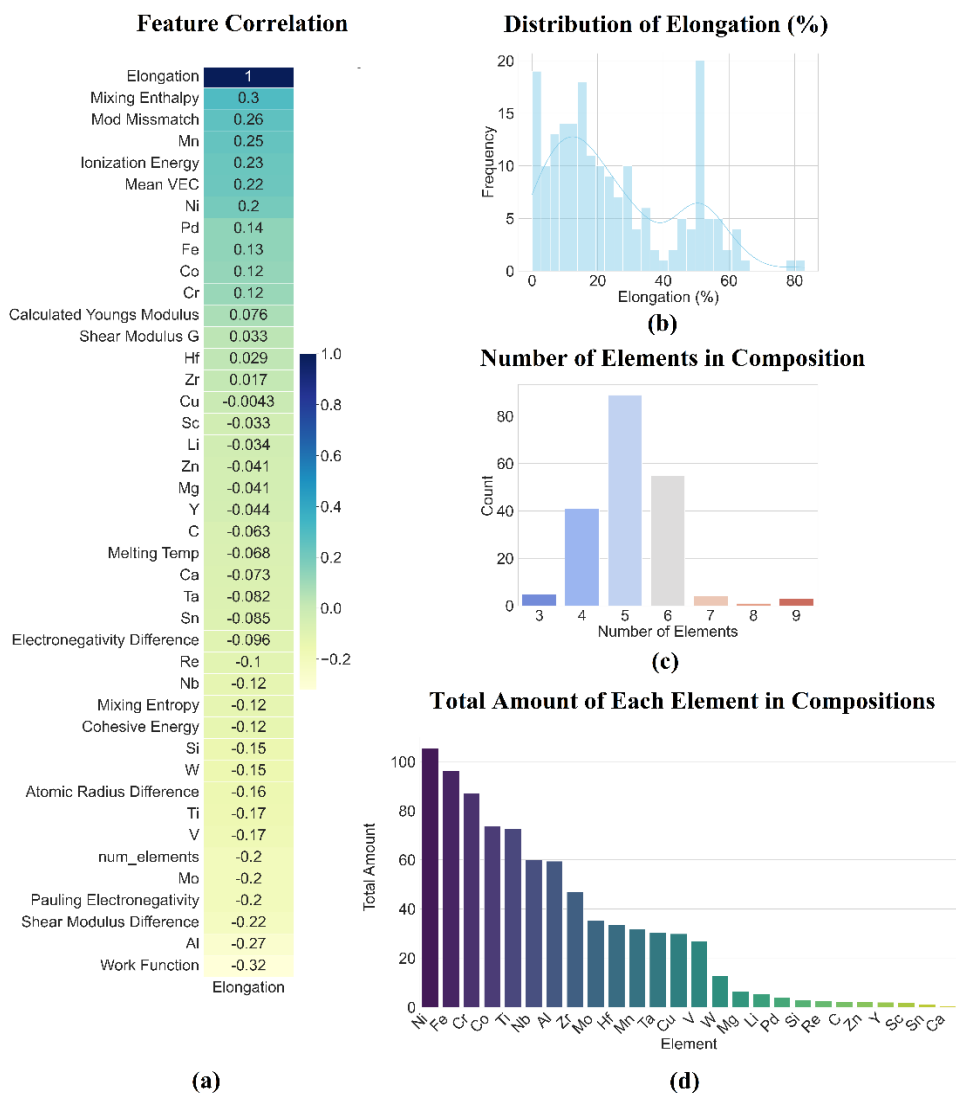

Figure 1. Dataset Analysis of the Elongation Target Property

**Ultimate Tensile Strength (UTS) Fine Tuning Dataset**

Similarly, the bar chart in Figure 2.d illustrates the total occurrence of each element in the dataset, showing that Ni, Fe, Ti, Co, and Cr are the most prevalent. Likewise, this dominance of specific elements could introduce bias into the model's predictions, making it more accurate for alloys with these common elements. The histogram (Figure 2.b) displaying the UTS distribution shows a considerable spread, with UTS values varying broadly across the dataset. Despite this wide spread, the distribution is more uniform than elongation, suggesting that a regression model might perform better on UTS predictions compared to elongation, given that UTS often has a more direct relationship with elemental composition. In heatmap (Figure 2.a) appears that the most correlated features include Shear Modulus Difference, Mixing Entropy, and Electronegativity Difference. Several other features, like Cohesive Energy and Melting Temperature, also contribute moderately, while negative correlations, such as with modulus mismatch and mean VEC, indicate that not all factors have a straightforward impact on UTS. These correlations suggest that the dataset captures meaningful relationships, but the weaker features may hinder the model's ability to make precise predictions. The composition of alloys based on the number of elements, as shown in Figure 2.c, reveals that most alloys contain 5 to 6 elements. This skew is common in HEA studies but could present a challenge, as compositions with 7 or more elements are underrepresented. In conclusion, despite the small dataset and compositional imbalances, the regression task for UTS predictions may result in better performance than elongation, as the correlations between features and UTS are stronger and more direct.

**Outlier Detection**

The two datasets were analysed to identify to what extend outliers may be contained in the dataset. The analysis has been done using the Z-scores approach. For a given dataset, the Z-score of an individual data point $x_i$ is calculated using the following formula:

$$Z_i = \frac{x_i - \mu}{\sigma}$$

where: $x_i$ = individual data point, $\mu$ = mean (average) of the dataset and $\sigma$ = standard deviation of the dataset.

This begins by calculating the mean of the target property measurements. The mean serves as a central reference point, around which most data points are expected to cluster. Alongside the mean, the standard deviation is determined to assess how much the target property values vary or spread out from this average. A smaller standard deviation indicates that the data points are tightly grouped around the mean, suggesting consistency within the dataset. Conversely, a larger standard deviation implies greater variability, meaning the data points are more dispersed over a wider range of values. Once the mean and standard deviation are established, each individual measurement is standardized to determine its relative position within the overall data distribution.

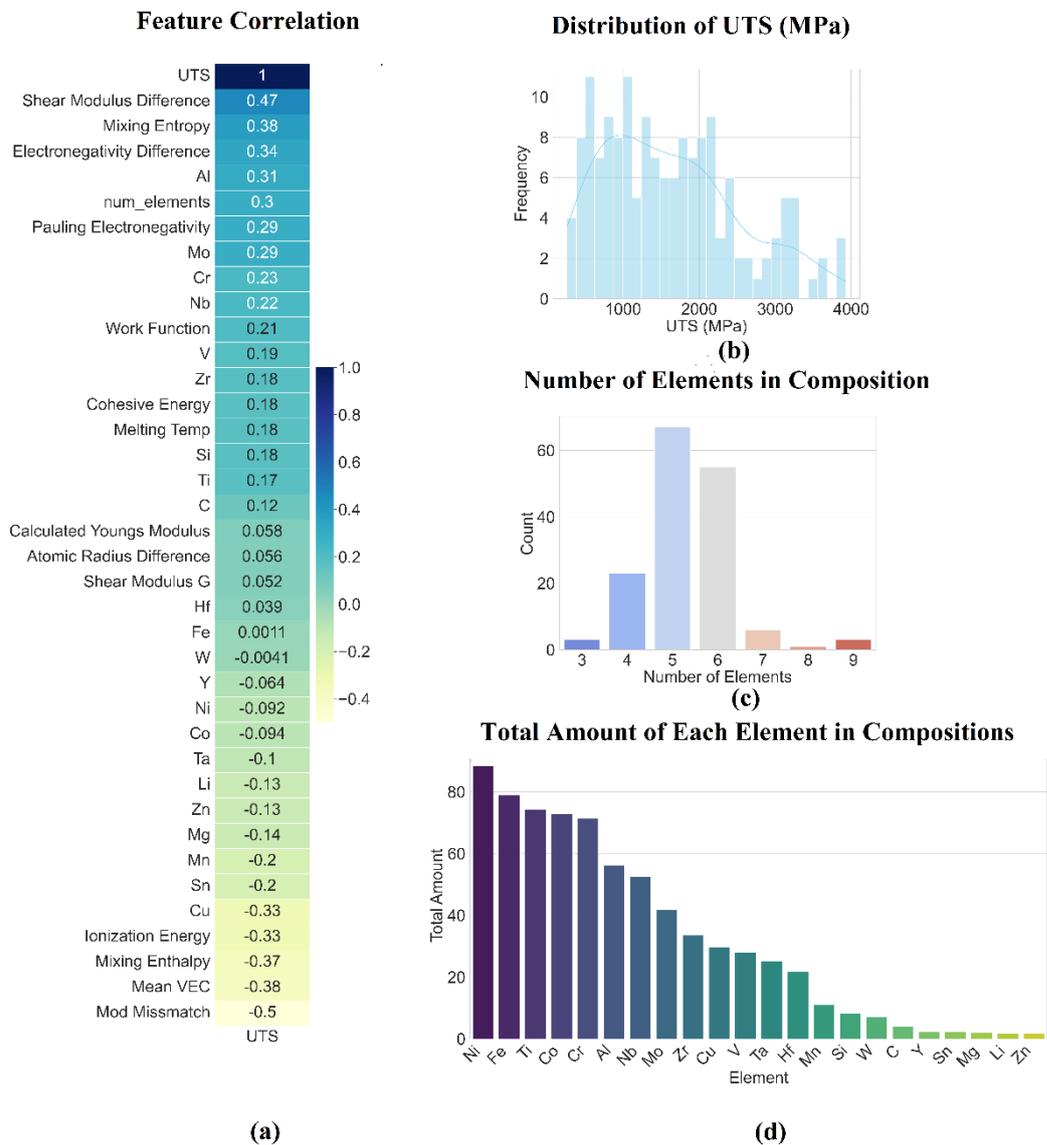

Figure 2. Dataset Analysis of the UTS Target Property

This standardization process results in the Z-score for each data point. A positive Z-score indicates that the value is above the mean, while a negative Z-score signifies that it is below the mean. This standardization transforms the data into a common scale, allowing for meaningful comparisons across different measurements. To identify outliers, the absolute value of each Z-score is considered, ensuring that both unusually high and unusually low values are accounted regardless of their direction. A common practice is to set a threshold for the Z-score, typically at 3. According to the empirical rule, in a normal distribution, approximately 99.7% of data points lie within three standard deviations from the mean. Therefore, any data point with an absolute Z-score exceeding this threshold is regarded as rare and potentially erroneous, qualifying it as an outlier.

In Figure 3.a, elongation values decrease rapidly from the highest point, with most data following a smooth downward trend. A single outlier, marked in red, stands out at the beginning of the plot, showing a much higher value compared to the other points. This deviation is also captured by the Z-score approach, as it significantly exceeds the threshold. In Figure 3.b, the data follows a rising trend with most points forming a nearly linear pattern until the higher values, where the points sharply increase. Although no outliers are explicitly marked, data points at the upper end of the scale some deviation from the mean are expected to form potential outliers. Overall, the Z-scores method identifies some extreme values but not significant outliers were detected to denote extensive rare events, or anomalies.

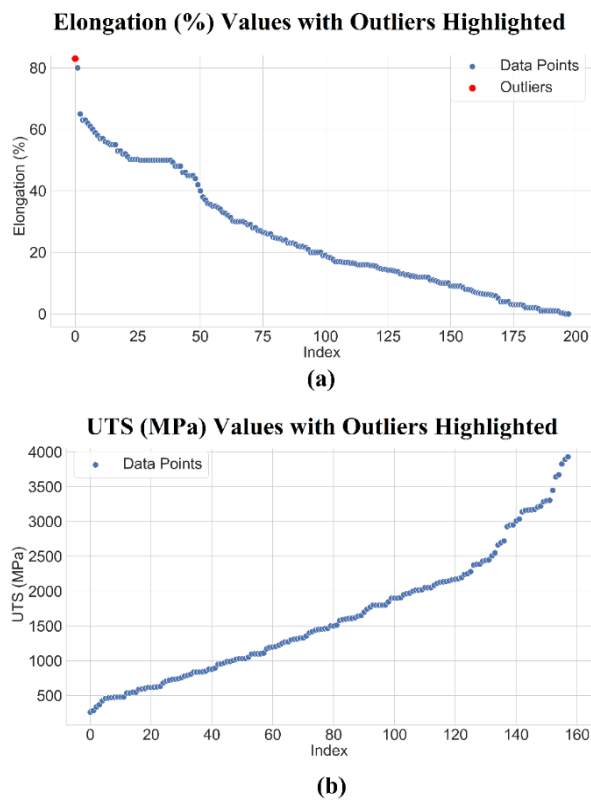

Figure 3. Z-score for both datasets (a. Elongation and b. UTS)

**Feature Engineering**

The feature engineering process begins by parsing chemical formulas to extract individual elements and their respective fractions within each alloy. Using regular expressions, the script identifies each element and determines its proportion, ensuring accurate representation of the alloy's composition. Once the elemental composition is established, the script computes a series of thermodynamic properties that are pivotal for understanding and predicting the behaviour of materials. The formulas for calculating the properties are included in Table 1. These properties among others, include Mean Valence Electron Concentration (Mean VEC), which averages the number of valence electrons per atom and influences electrical and magnetic characteristics. Electronegativity Difference assesses the disparity in electron-attracting abilities among elements, affecting bond strength and corrosion resistance. Atomic Radius Difference evaluates variations in atomic sizes, impacting lattice strain and mechanical strength.

The script further estimates the Young's Modulus, a measure of the alloy's stiffness and resistance to elastic deformation, and Shear Modulus, which quantifies resistance to shear forces and overall mechanical robustness. Modulus Mismatch and Shear Modulus Difference highlight disparities in mechanical properties among constituent elements, indicating potential internal stresses that could affect durability and performance. Mixing Enthalpy and Mixing Entropy provide insights into the thermodynamic stability and disorder introduced during alloy formation, influencing phase formation and material homogeneity. Additional features such as Electron Work Function measure the energy required to remove electrons from the material's surface, impacting electrical conductivity and catalytic activity. Melting Temperature predicts the thermal stability and suitability of alloys for high-temperature applications, while Cohesive Energy reflects the bond strength within the alloy, influencing hardness and durability. Average Ionization Energy assesses the energy required to ionize atoms, affecting electrical and thermal properties. The script also includes Electronegativity Difference using the Pauling scale and Latent Heat, which represents the energy involved in phase transitions crucial for processing and thermal management. By integrating these features into the dataset, the approach provides a comprehensive and multifaceted view of each alloy's properties. This enriched dataset equips machine learning models with the necessary information to predict complex material behaviours.

| Property | Symbol | Formula | Units | Description |
|---|---|---|---|---|
| Mean VEC | VEC | $\sum_i x_i \cdot VEC_i$ | electrons/atom | Average number of valence electrons per atom. |
| Electronegativity Difference | $\delta_x$ | $\sqrt{\sum_i x_i (X_i - \bar{X})^2}$ | None | Standard deviation of electronegativity. |
| Atomic Radius Difference | $\delta_r$ | $\sum_i x_i \left(\frac{r_i - \bar{r}}{\bar{r}}\right)^2 \times 100$ | % | Percentage difference in atomic radius. |
| Calculated Young's Modulus | $E$ | $\sum_i x_i \cdot E_i$ | GPa | Weighted average of Young's modulus. |
| Mixing Enthalpy | $\Delta H_{mix}$ | $\sum_i \sum_{j \neq i} x_i x_j \Delta H_{ij}$ | kJ/mol | Enthalpy change during mixing of components. |
| Mixing Entropy | $\Delta S_{mix}$ | $-R \sum_i x_i \ln x_i$ | J/(mol·K) | Entropy change during mixing of components. |
| Work Function | $\varphi$ | $\sum_i x_i \cdot \varphi_i$ | eV | Average electron work function of the alloy. |
| Shear Modulus | $G$ | $\sum_i x_i \cdot G_i$ | GPa | Weighted average of shear modulus. |
| Modulus Mismatch | $\Delta M$ | $\sum_i x_i \left(2 \frac{(G_i - \bar{G})}{G_i + \bar{G}}\right)^2$ | None | Mismatch in modulus across alloy components. |
| Shear Modulus Difference | $\Delta G$ | $\sum_i x_i \left(1 - \frac{G_i}{\bar{G}}\right)^2$ | None | Difference in shear modulus relative to the mean. |
| Melting Temperature | $T_m$ | $\sum_i x_i \cdot T_i$ | K | Weighted average melting point of the alloy. |
| Cohesive Energy | $E_{coh}$ | $\sum_i x_i \cdot E_{coh,i}$ | eV/atom | Average energy needed to break atomic bonds. |
| Ionization Energy | $IE$ | $\sum_i x_i \cdot IE_i$ | eV | Average first ionization energy. |
| Pauling Electronegativity | $\Delta X_p$ | $\sum_i \sum_{j \neq i} x_i x_j (X_{P,i} - X_{P,j})^2$ | None | Measure of electronegativity difference using Pauling scale |

$x_i$ : is the atomic fraction of component $i$, $X_i$ : Electronegativity of component $i$, $r_i$ : atomic radius of component $i$, $\bar{X}$ and $\bar{r}$ are the average electronegativity and atomic radius of the alloy, $E_i$, $G_i$, $T_i$, $\varphi_i$, $E_{coh,i}$, $IE_i$ are the Young's modulus, shear modulus, melting temperature, work function, cohesive energy, and ionization energy of component $i$, respectively. $\Delta H_{ij}$ is the mixing enthalpy between components $i$, $j$ and $R$ is the universal gas constant.

Table 1. Thermodynamic Properties used as Input Features

**Generation of Pre-training Dataset**

In the preparation of the pre-training dataset for predicting material properties, a systematic approach was employed to generate and curate a diverse set of alloy compositions. Initially, a comprehensive list of metallic elements was established, each assigned specific weights to influence their selection probability. This weighting mechanism ensured a balanced and realistic distribution of elements across the generated alloys, reflecting their natural abundance and relevance in material science. To achieve this diversity, compositions were categorized into equimolar and non-equimolar types. Equimolar alloys featured elements in equal proportions, promoting uniformity and simplifying the analysis of elemental interactions. In contrast, non-equimolar alloys incorporated unequal proportions of elements, introducing variability and complexity that better mimic real-world materials. Uniqueness of each alloy composition was a critical consideration. By ensuring that no element was duplicated within a single alloy, the dataset maintained chemical validity and prevented redundancy. This was achieved through careful selection processes that randomly chose elements based on their weighted probabilities while enforcing the constraint of unique elemental presence within each composition.

The distribution of elements within the dataset was rigorously validated to confirm adherence to the intended frequencies and proportions. Visualization techniques, were employed to assess the occurrence of each element, ensuring that the dataset accurately represented the desired elemental distribution. Additionally, comprehensive checks were conducted to identify and eliminate any duplicate entries that could potentially skew the dataset's integrity. Further refinement involved filtering out any compositions with zero values in critical feature columns. This step was essential to prevent the introduction of misleading or incomplete data points that could adversely affect the pre-training process. The final dataset was enriched with a wide array of the thermodynamic properties (features) as presented in Table 1.

To further assess the pre-training dataset quality, we applied a t-distributed Stochastic Neighbor Embedding (t-SNE) approach (Figure 4). This dimensionality reduction tool transforms high-dimensional data into a two-dimensional space, allowing for the visualization of complex relationships and structures within the data. When analysing pre-training and fine-tuning datasets, t-SNE plots may offer valuable insights into how these datasets interact within a shared feature space. By representing each data point in two dimensions, t-SNE helps illustrate the similarity and clustering of data from different sources, such as pre-training datasets and specific fine-tuning tasks like Elongation and UTS.

In this work we utilise three pre-training datasets with varying sizes: 6K, 75K, and 150K entries. As the size of the pre-training dataset increases to 150K entries, the t-SNE plot reveals a denser and more compact feature space. This density indicates that the model has captured a broader and more nuanced array of patterns and representations from the extensive pre-training data. Consequently, the fine-tuning datasets for Elongation and UTS show greater overlap with these dense pre-training clusters. This enhanced overlap suggests that the model can effectively leverage the rich, generalized features learned during pre-training, leading to improved performance in downstream regression tasks. Moreover, a denser feature space facilitates better generalization, enabling the model to perform reliably on unseen data that falls within the comprehensive feature landscape established by the larger pre-training dataset.

However, while t-SNE plots are powerful for visualizing data relationships, they come with certain limitations. The stochastic nature of t-SNE means that results can vary between runs, potentially affecting reproducibility. Additionally, t-SNE is primarily effective at preserving local structures, which means that the global relationships between clusters might be distorted, leading to possible misinterpretations of the data's true structure. Computational intensity is another concern, especially with very large datasets, as t-SNE can be resource-demanding and time-consuming. Furthermore, the visualization outcome is sensitive to hyperparameters like perplexity and learning rate, requiring careful tuning to avoid misleading representations. Despite these limitations, when used appropriately, t-SNE plots remain a valuable tool for assessing the alignment and overlap between pre-training and fine-tuning datasets, providing critical insights that can guide model improvements and enhance overall performance.

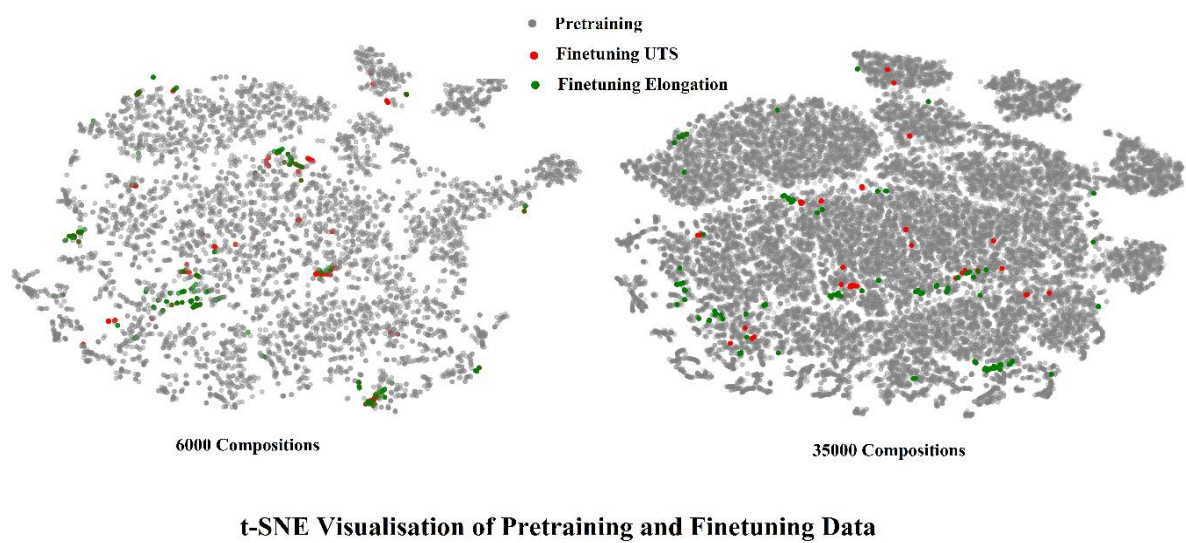

Figure 4. t-SNE representation of pre-training and fine-tuning data

**Pre-Training the Transformer for Downstream Tasks**

The pre-training phase of our study leverages a BERT-based Masked Language Model (MLM) [44] explicitly tailored for chemical composition data integrated with numerical features. The process is shown schematically in Figure 5. BERT is selected for material predictions after pre-training and fine-tuning because effectively understands complex contexts through its bidirectional processing, which is particularly important for analyzing elemental sequences where the relationship between elements depends on their surrounding context. By leveraging extensive pre-trained knowledge for transfer learning, BERT enhances its adaptability to various prediction tasks. Its ability to handle structured and intricate input representations, such as chemical formulas and elemental arrangements, combined with state-of-the-art performance and rich feature extraction, makes it highly suitable for material science applications.

This process begins with a custom tokenization strategy designed to accurately parse and represent chemical formulas. For instance, a composition like "Co1.2 Fe0.8 Ni1" is processed by extracting elemental fraction pairs (Co, 1.2), (Fe, 0.8), and (Ni, 1). These pairs are then sorted alphabetically and concatenated to form tokens such as "Co1.2 Fe0.8 Ni1". Mathematically, each token $T_i$ is constructed as $T_i = E_i \circ f_i$. where $E_i$ is the element symbol and $f_i$ its fraction. Once tokenized, these chemical compositions are combined with additional numerical features from the dataset. Numerical columns are converted to strings and appended to the tokenized text, resulting in a comprehensive input like "Co1.2 Fe0.8 Ni1 300 500", where the numbers represent specific thermodynamic properties.

The combined text data is then split into training and validation sets (80-20%) to facilitate effective model evaluation. Utilizing the BERT tokenizer, each text sequence undergoes further tokenization, padding, and truncation to ensure uniform input lengths. During this stage, the Masked Language Modeling (MLM) objective is applied by randomly masking a subset of tokens within the input. For example, "Co1.2 Fe0.8 Ni1 300 500" might become "[MASK] Fe0.8 Ni1 300 500". The MLM head, comprising a linear layer, predicts the masked tokens based on their surrounding context. Specifically, the MLM head maps the final hidden states corresponding to the masked token positions to a probability distribution over the tokenizer's vocabulary:

$$\hat{T}_i = \text{softmax}(W_{\text{mlm}} H_i + b_{\text{mlm}})$$

where $H_i$ is the hidden state for the masked token, and $W_{\text{mlm}}$ and $b_{\text{mlm}}$ are the weights and biases of the MLM head. The model is trained to minimize the cross-entropy loss function:

$$\mathcal{L}_{\text{MLM}} = -\sum_{i \in \mathcal{M}} \log P(T_i \mid \text{Context})$$

where $\mathcal{M}$ represents the set of masked token positions, and $P(T_i \mid \text{Context})$ is the predicted probability of the true token given its surrounding context. Through this masked token prediction task, the model learns to generate contextualized embeddings that capture the intricate relationships between different elements within a composition and their associated thermodynamic properties. For instance, the model internalizes patterns such as how varying fractions of Fe might correlate with specific material thermodynamic properties like VEC or δ. A custom callback monitors validation loss during training, ensuring that the best-performing model is saved for subsequent fine-tuning. Upon completion, the pre-trained model encapsulates a deep understanding of chemical compositions, providing a robust foundation for downstream regression tasks aimed at predicting material properties based on complex chemical data. This meticulous pretraining approach, combining domain-specific tokenization with the powerful contextual learning capabilities of BERT, establishes a solid groundwork for accurately modeling and predicting thermodynamic properties from chemical compositions.

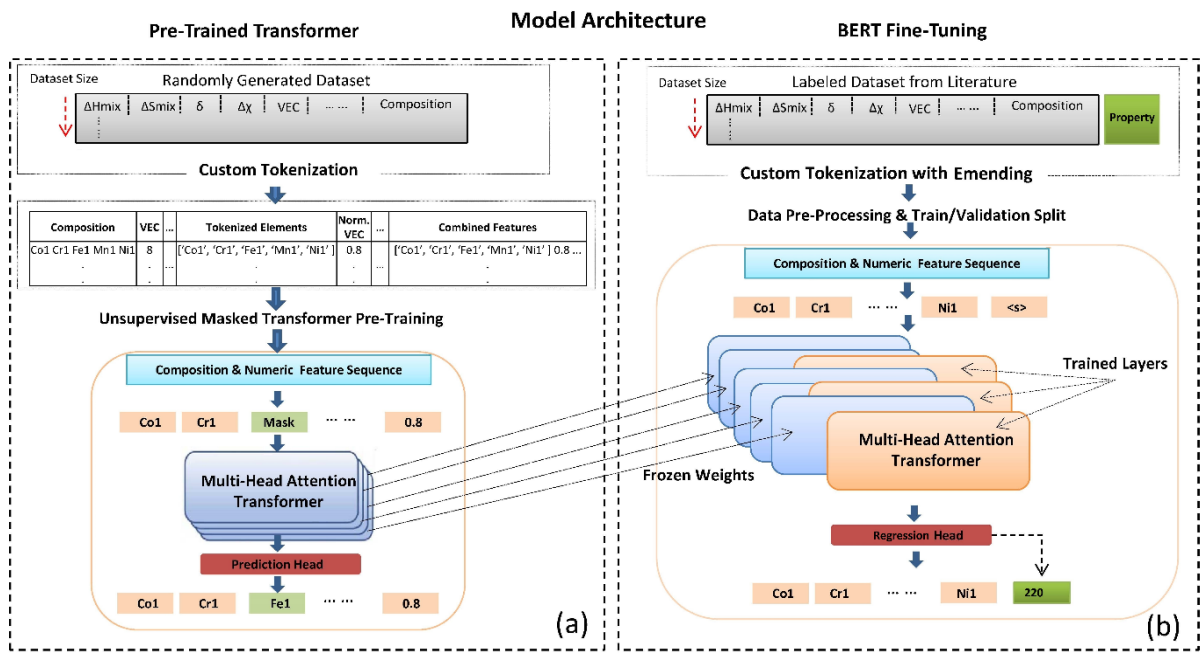

Figure 5: a) Transformer Pre-Training and b) Fine-Tuning

**Fine-Tuning Process of BERT Model for Regression**

The fine-tuning phase builds upon the pre-trained BERT-based MLM [44] to adapt it for regression tasks aimed at predicting thermodynamic properties from chemical compositions and numerical features. The process is shown schematically in Figure 5.b

This process begins by loading the pre-tokenized dataset, which comprises chemical compositions and their associated numerical attributes. To ensure robust evaluation and to mitigate overfitting, we employ $K$-fold cross-validation ($K = 5$). For each fold, the dataset is partitioned into training and validation subsets (80-20%). Within each training subset, numerical features are normalized using the Standard Scaler, transforming each feature $x_j$ into a standardized form:

$$\tilde{x}_j = \frac{x_j - \mu_{x_j}}{\sigma_{x_j}}$$

where $\mu_{x_j}$ and $\sigma_{x_j}$ represent the mean and standard deviation of feature $x_j$ within the training set. This normalization ensures that all numerical features contribute equally to the model's learning process. To prevent data leaks the scaler is applied after data split into training and validation and not before.

The normalized numerical features are then concatenated with the tokenized chemical compositions to form comprehensive input sequences, such as "Co1.2 Fe0.8 Ni1.0 7 400 ", where the numbers correspond to specific thermodynamic and target properties like UTS and elongation. Utilizing the BERT tokenizer, each combined text sequence undergoes further tokenization, padding, and truncation to a uniform length of 512 tokens, ensuring consistency across inputs. These tokenized sequences are organized into custom PyTorch datasets, facilitating efficient data handling during training.

Central to the fine-tuning process is the incorporation of the Multi-Head Self-Attention mechanism within the BERT architecture. For each token in the input sequence, the model computes three distinct vectors: Query (**Q**), Key (**K**), and Value (**V**). These vectors are derived through learned linear transformations of the token's hidden state:

$$\mathbf{Q}_i = \mathbf{W}_Q \mathbf{H}_i, \quad \mathbf{K}_i = \mathbf{W}_K \mathbf{H}_i, \quad \mathbf{V}_i = \mathbf{W}_V \mathbf{H}_i$$

where $\mathbf{H}_i$ is the hidden state of the $i$-th token, and $\mathbf{W}_Q, \mathbf{W}_K, \mathbf{W}_V$ are learned weight matrices. The self-attention mechanism calculates attention scores by taking the dot product of the Query vector of one token with the Key vectors of all tokens in the sequence, scaling them by the square root of the dimensionality $\left(\sqrt{d_k}\right)$, and applying a softmax function to obtain attention weights:

$$\alpha_{ij} = \frac{\exp\left(\frac{\mathbf{Q}_i \cdot \mathbf{K}_j^\top}{\sqrt{d_k}}\right)}{\sum_{k=1}^{L} \exp\left(\frac{\mathbf{Q}_i \cdot \mathbf{K}_k^\top}{\sqrt{d_k}}\right)}$$

These attention weights are then used to compute a weighted sum of the Value vectors, producing an output that captures contextual information from the entire sequence. By employing multiple attention heads, the model can attend to different aspects of the input simultaneously, enhancing its ability to capture complex relationships between elements in the chemical compositions and their corresponding thermodynamic and mechanical properties.

Following the self-attention layers, the model includes a regression head-a linear layer that maps the [CLS] token's final hidden state to a single scalar value:

$$\hat{y} = \mathbf{W}_{\text{reg}}^\top \mathbf{h}_{\text{CLS}} + b_{\text{reg}}$$

where $\mathbf{h}_{\text{CLS}}$ is the hidden state of the [CLS] token, $\mathbf{W}_{\text{reg}}$ is the weight vector, and $b_{\text{reg}}$ is the bias term. This regression head enables the model to output continuous predictions corresponding to the thermodynamic properties of interest.

The optimization process utilizes the AdamW optimizer that incorporates weight decay for regularization. Parameters from specific Transformer layers, particularly the last three layers, are subjected to weight decay ($\lambda = 0.02$), while others are exempt. This selective regularization helps prevent overfitting by penalizing large weights in critical parts of the model. The learning rate is set to $\eta = 6 \times 10^{-5}$, and a linear learning rate scheduler with warm-up steps is applied to facilitate smooth convergence during training.

Training is managed using the Trainer API, which oversees the optimization loop, handles gradient calculations, and applies the defined learning rate schedule. The Mean Squared Error (MSE), Mean Absolute Error (MAE), and the coefficient of determination ($R^2$) are employed as evaluation metrics to assess the model's performance:

$$\text{MSE} = \frac{1}{N}\sum_{i=1}^{N}(\hat{y}_i - y_i)^2, \ \text{MAE} = \frac{1}{N}\sum_{i=1}^{N}|\hat{y}_i - y_i|, \ R^2 = 1 - \frac{\sum_{i=1}^{N}(\hat{y}_i - y_i)^2}{\sum_{i=1}^{N}(y_i - \bar{y})^2}$$

where $\hat{y}_i$ and $y_i$ denote the predicted and true values, respectively, and $\bar{y}$ is the mean of the true values.

A custom callback monitors validation loss throughout the training process, ensuring that the model with the lowest validation loss is saved for each fold. This best-performing model is then used for evaluation and further analysis. Through the integration of the Multi-Head Self-Attention mechanism, the model effectively learns to generate contextualized embeddings that capture the intricate relationships between different elements within a chemical composition and their influence on thermodynamic properties. For example, the model may recognize that higher fractions of Ni correlate with increased elongation values, adjusting its internal representations accordingly.

After completing all folds, the performance metrics are aggregated to provide an overall assessment of the model's predictive capabilities. Residual analyses, including scatter plots of actual versus predicted values and distribution plots of residuals, are conducted to visualize and interpret the model's accuracy and potential biases. The fine-tuning process results in a robust model capable of accurately predicting material properties based on complex chemical compositions and numerical data, thereby offering valuable insights for materials science applications.

**Multiple Regressor Training**

In this phase of our study, we focus on training and evaluating a suite of regression models to predict mechanical macroscopic properties, specifically Elongation and UTS, based on chemical compositions and associated thermodynamic features (Table 1). The approach employs various machine learning algorithms to ascertain the most effective model for our predictive task, ensuring comprehensive coverage of different modeling paradigms.

The process commences with data loading, where the relevant dataset is imported using pandas. The dataset contains a 'composition' column, representing chemical formulas (Feature), 14 thermodynamic numerical property columns (Features) and a 'target property' column. To prepare the data for regression, non-numeric columns, particularly 'composition', are processed to extract elemental fractions. A custom function leverages regular expressions to parse each chemical composition string, extracting element-symbol and fraction pairs. For example, a composition like "Co1.2 Fe0.8 Ni1" is transformed into a dictionary: $\mathcal{E}(S) = \{(Co, 1.2), (Fe, 0.8), (Ni, 1.0)\}$. These elemental fractions are then organized into a DataFrame, with missing elements filled with zeros to maintain consistent feature dimensions across samples.

Subsequently, the numerical features are combined with the extracted elemental fractions to form the feature matrix **X**, while the target vector **y** comprises the mechanical property values that we want to train the model to predict. To ensure that all features contribute uniformly to the learning process, we apply standard normalization using the Standard Scaler in a similar way as for the language model approach. The same K-Fold cross-validation with K=5 is also employed for the regression models ensuring that each model is evaluated on diverse data splits, enhancing the reliability of performance metrics.

The Gaussian Process Regressor models the target variable $y$ as a realization of a Gaussian process, characterized by a mean function $m(\mathbf{x})$ and a covariance function $k(\mathbf{x}, \mathbf{x}') : y(\mathbf{x}) \sim \mathcal{GP}(m(\mathbf{x}), k(\mathbf{x}, \mathbf{x}'))$. Given training data $\mathbf{X}_{\text{train}}$ and $\mathbf{y}_{\text{train}}$, the GPR predicts the distribution of $y_*$ at a new input $\mathbf{x}_*$ as: $y_* \mid \mathbf{x}_*, \mathbf{X}_{\text{train}}, \mathbf{y}_{\text{train}} \sim \mathcal{N}(\mu_*, \sigma_*^2)$, where:

$$\mu_* = \mathbf{k}_*^\top (\mathbf{K} + \sigma_n^2 \mathbf{I})^{-1} \mathbf{y}_{\text{train}}$$
$$\sigma_*^2 = k(\mathbf{x}_*, \mathbf{x}_*) - \mathbf{k}_*^\top (\mathbf{K} + \sigma_n^2 \mathbf{I})^{-1} \mathbf{k}_*$$

Here, **K** is the covariance matrix computed from the training inputs, $\mathbf{k}_*$ is the covariance vector between the new input and training inputs, and $\sigma_n^2$ represents the noise variance. Complete list of symbol explanations is included in Appendix A.

The Random Forest Regressor operates by constructing an ensemble of decision trees during training. Each tree $T_i$ is trained on a bootstrap sample of the training data and makes individual predictions $T_i(\mathbf{x})$. The final prediction is the average of all tree predictions:

$$\hat{y}(\mathbf{x}) = \frac{1}{N} \sum_{i=1}^{N} T_i(\mathbf{x})$$

This aggregation reduces variance and enhances generalization by leveraging the wisdom of multiple trees. The Decision Tree Regressor partitions the feature space into hierarchical regions based on feature thresholds, forming a tree-like structure. Each leaf node represents a region where the prediction is the mean of the target values of the training samples within that region:

$$\hat{y}(\mathbf{x}) = \frac{1}{|\mathcal{L}(\mathbf{x})|} \sum_{i \in \mathcal{L}(\mathbf{x})} y_i$$

where $\mathcal{L}(\mathbf{x})$ denotes the set of training samples falling into the same leaf node as the input $\mathbf{x}$.

The Gradient Boosting Regressor builds an additive model by sequentially training decision trees to minimize a specified loss function. At each iteration $m$, a new tree $T_m$ is trained on the residuals from the previous ensemble:

$$\hat{y}^{(m)}(\mathbf{x}) = \hat{y}^{(m-1)}(\mathbf{x}) + \eta T_m(\mathbf{x})$$

where $\eta$ is the learning rate. The residuals are computed as:

$$r_i^{(m)} = y_i - \hat{y}^{(m-1)}(\mathbf{x}_i)$$

This approach allows the model to focus on correcting the errors of the prior ensemble, leading to improved performance over iterations. The K-Nearest Neighbors Regressor predicts the target value for a new input $\mathbf{x}$ by averaging the target values of its $K$ nearest neighbors in the feature space:

$$\hat{y}(\mathbf{x}) = \frac{1}{K} \sum_{i=1}^{K} y_i$$

where the nearest neighbors are determined based on a distance metric, typically Euclidean distance. This non-parametric method relies on the assumption that similar inputs have similar target values.

**Results and Discussions**

In this study, all models utilized the same set of features as depicted in Table 1. Hyperparameter tuning was not performed for the models in this work. The decision to omit hyperparameter optimization was made to maintain consistency across all models and to focus on evaluating their baseline performances under default settings. By using the default hyperparameters provided by the relevant machine learning libraries (scikit-learn, Pytorch Transformers), we aimed to reduce computational complexity and avoid potential data leakage that could arise from extensive hyperparameter searches, especially given the limited size of our dataset. Additionally, all models were trained on the same uncleaned dataset to ensure that the comparisons were fair and solely attributed to the models' inherent capabilities rather than pre-processing differences.

Model comparisons were made against appropriate baselines by evaluating each model's performance using standard metrics such as R-squared (R²), Mean Squared Error (MSE) and Mean Average Error (MAE). To enhance reproducibility and usability, the dataset used for training and evaluating the models is openly available alongside the codebase at [https://github.com/SPS-Coatings/Language-Model-for-HEA]. The training runs were conducted on a computing infrastructure comprising two NVidia A5000 GPUs with 24Gb RAM each running on Windows OS, using Python 3.9 with all dependencies documented in the github link alongside the requirements txt file. The total time taken to pre-train the transformer with 150K entries was approximately 48 hours. Fine-tuning was much faster requiring approximately 10 minutes for each dataset and model parameters. Finally, no specific benchmark frameworks were used as the objective of this work is not to present a model with higher accuracy than others reported in the literature, but rather to demonstrate the language model advantages compared to other models when using same datasets, pre-processing, feature engineering, and training frameworks.

**Model Evaluation**

An extensive analysis has been performed to evaluate the performance of pre-trained Transformer models in predicting material properties, with emphasis on elongation and UTS, in comparison with traditional machine learning models. The results, presented in Tables 2 through 6, consistently demonstrate the superior predictive capabilities of the pre-trained Transformer models. This superiority is attributed to the Transformer's advanced architecture, effective pre-training strategies, and its ability to learn complex representations from large datasets, even when only elemental compositions are provided as input without additional engineered features.

Table 2 shows the performance of various models on elongation prediction. The pre-trained Transformer achieved the lowest mean squared error (MSE) of 0.428 and mean absolute error (MAE) of 0.449, along with the highest coefficient of determination (R²) of 0.561. In contrast, the Transformer without pre-training recorded a higher mean MSE of 0.457 and a lower mean R² of 0.531. Traditional models like Random Forest and Gaussian Process performed reasonably well but did not surpass the pre-trained Transformer, with mean MSE values of 0.452 and 0.470, respectively. The significant performance gap between the pre-trained and non-pre-trained Transformers underscores the critical role of pre-training in enhancing the model's ability to capture the underlying patterns related to elongation.

| Table 2 | Performance of Transformer Model and Baseline Models on Elongation (Test Set) | | | | | |
|---|---|---|---|---|---|---|
| Model | Mean MSE ($\downarrow$) | Best k-fold MSE ($\downarrow$) | Mean MAE ($\downarrow$) | Best k-fold MAE ($\downarrow$) | Mean $R^2$ ($\uparrow$) | Best k-fold $R^2$ ($\uparrow$) |
| Gausian Process | 0.470 | 0.389 | 0.524 | 0.438 | 0.522 | 0.60 |
| Random Forest | 0.452 | 0.358 | 0.504 | 0.431 | 0.530 | 0.67 |
| K-NN | 0.623 | 0.509 | 0.618 | 0.566 | 0.356 | 0.46 |
| Descition Trees | 0.728 | 0.515 | 0.570 | 0.504 | 0.245 | 0.49 |
| Gradient Boosting | 0.538 | 0.416 | 0.539 | 0.454 | 0.442 | 0.59 |
| **Transformer ( Pre-trained)** | **0.428** | 0.349 | **0.449** | 0.392 | **0.561** | 0.67 |
| Transformer ( Not Pre-trained) | 0.457 | 0.338 | 0.498 | 0.408 | 0.531 | 0.65 |

Metrics for scaled values. The bold indicate the best mean results in terms of the metrics used with 5 fold validation.
The transformer models were trained across all layers using all features as model inputs

Similarly, Table 3 presents the models' performance on UTS prediction, where the pre-trained Transformer again outperformed all other models. It achieved the lowest mean MSE of 0.213 and the highest mean R² of 0.785. The non-pre-trained Transformer exhibited a substantial decline in performance, with a mean MSE of 0.336 and a mean R² of 0.619. Baseline models such as Gradient Boosting and Random Forest showed competitive performance but still fell short of the pre-trained Transformer's accuracy. These findings highlight the effectiveness of pre-training in equipping the Transformer with a superior understanding of the relationships between material compositions and properties.

| Table 3 | Performance of Transformer Model and Baseline Models on UTS (Test Set) | | | | | |
|---|---|---|---|---|---|---|
| Model | Mean MSE ($\downarrow$) | Best k-fold MSE ($\downarrow$) | Mean MAE ($\downarrow$) | Best k-fold MAE ($\downarrow$) | Mean $R^2$ ($\uparrow$) | Best k-fold $R^2$ ($\uparrow$) |
| Gausian Process | 0.282 | 0.171 | 0.370 | 0.286 | 0.719 | 0.78 |
| Random Forest | 0.251 | 0.146 | 0.354 | 0.287 | 0.747 | 0.86 |
| K-NN | 0.375 | 0.287 | 0.434 | 0.372 | 0.621 | 0.70 |
| Descition Trees | 0.571 | 0.336 | 0.505 | 0.370 | 0.390 | 0.65 |
| Gradient Boosting | 0.226 | 0.085 | 0.341 | 0.211 | 0.764 | 0.86 |
| **Transformer ( Pre-trained)** | **0.213** | 0.153 | **0.336** | 0.291 | **0.785** | 0.79 |
| Transformer ( Not Pre-trained) | 0.336 | 0.256 | 0.420 | 0.353 | 0.619 | 0.73 |

Metrics for scaled values. The bold indicate the best mean results in terms of the metrics used with 5 fold validation.

Table 4 delves into the impact of the pre-training dataset size on the Transformer's performance. As the number of compositions used for pre-training increased from 6,000 to 150,000, there was a notable improvement in the model's predictive capabilities, particularly for UTS. The mean MSE for UTS decreased from 0.382 to 0.251, and the mean R² increased from 0.617 to 0.754. This trend indicates that a larger pre-training dataset enables the Transformer to learn more comprehensive and robust representations of material behaviors.

For elongation, the improvements were more modest but still evident, suggesting that elongation may be influenced by factors that require both extensive data exposure and fine-tuning strategies.

| Table 4 | Effect of Pre-Training Dataset Size on the Fine Tuned Transformer Model Performance (Test Set) | | | | | |
|---|---|---|---|---|---|---|
| Pre-Trained Model | Mean MSE Elongation ($\downarrow$) | Mean MSE UTS ($\downarrow$) | Mean MAE Elongation ($\downarrow$) | Mean MAE UTS ($\downarrow$) | Mean $R^2$ Elongation ($\uparrow$) | Mean $R^2$ UTS ($\uparrow$) |
| 6.000 Compositions | 0.434 | 0.382 | 0.498 | 0.448 | 0.553 | 0.617 |
| 75.000 Compositions | 0.439 | 0.312 | 0.478 | 0.414 | 0.551 | 0.685 |
| 150.000 Compositions | **0.428** | **0.251** | **0.449** | **0.349** | **0.561** | **0.754** |

Metrics for scaled values. The bold indicate the best mean results in terms of the metrics used with 5 fold validation.
The transformer models were trained across all layers using all features as model inputs

In Table 5, the effect of fine-tuning is explored for different numbers of attention layers within the Transformer. The results reveal that fine-tuning only the deeper layers, specifically layers 4, 6 8, 10 and 12, yielded the best performance for UTS prediction, with a mean MSE of 0.210 and a mean $R^2$ of 0.786. This suggests that higher-level abstractions captured in these layers are particularly pertinent to UTS. In contrast, fine-tuning all layers resulted in the best performance for elongation prediction, achieving a mean MSE of 0.428 and a mean $R^2$ of 0.561. This indicates that both low-level and high-level features are important for accurately predicting elongation, and that the contributions from all layers collectively enhance the model's performance for this property.

| Table 5 | Effect of Number of Trained Attention Layers on the Fine Tuned Transformer Model Performance (Test Set) | | | | | |
|---|---|---|---|---|---|---|
| Fine Tuning | Mean MSE Elongation ($\downarrow$) | Mean MSE UTS ($\downarrow$) | Mean MAE Elongation ($\downarrow$) | Mean MAE UTS ($\downarrow$) | Mean $R^2$ Elongation ($\uparrow$) | Mean $R^2$ UTS ($\uparrow$) |
| All Layers | **0.428** | 0.250 | **0.449** | 0.348 | **0.561** | 0.754 |
| Layers 11, 12 | 0.445 | 0.212 | 0.478 | **0.335** | 0.545 | 0.785 |
| Layers 10, 11, 12 | 0.471 | 0.237 | 0.499 | 0.344 | 0.518 | 0.767 |
| Layers 9, 10, 11, 12 | 0.445 | 0.218 | 0.475 | 0.349 | 0.540 | 0.781 |
| Layers 4, 6, 8, 10, 12 | 0.446 | **0.210** | 0.484 | 0.342 | 0.538 | **0.786** |

Metrics for scaled values. The bold indicate the best mean results in terms of the metrics used with 5 fold validation.
Layer-wise fine-tuning. The attention weights are kept frozen for the unselected layers. 150,000 Compositions Pre-trained Model with all input Features

Table 6 examines the Transformer's performance when only elemental compositions are used as input, without additional engineered features. Remarkably, the Transformer maintained superior performance over baseline models even in this scenario. For elongation prediction, the Transformer fine-tuned on layers 10 to 12 achieved a mean MSE of 0.436 and a mean $R^2$ of 0.554, outperforming the Random Forest and Gaussian Process models, which had mean MSE values of 0.484 and 0.510, respectively. For UTS prediction, the Transformer's performance remained robust, further emphasizing its ability to extract meaningful patterns directly from raw input data. The baseline models, on the other hand, exhibited reduced performance without engineered features, highlighting their reliance on such inputs for accurate predictions.

| Table 6 | Effect of Features on the Fine Tuned Transformer Model Performance (Test Set) | | | | | |
|---|---|---|---|---|---|---|
| Models trained with composition as input only | Mean MSE Elongation ($\downarrow$) | Mean MSE UTS ($\downarrow$) | Mean MAE Elongation ($\downarrow$) | Mean MAE UTS ($\downarrow$) | Mean $R^2$ Elongation ($\uparrow$) | Mean $R^2$ UTS ($\uparrow$) |
| All Layers No Features | 0.453 | 0.250 | 0.492 | 0.361 | 0.532 | 0.755 |
| Layers 11, 12 No features | 0.462 | 0.258 | 0.497 | 0.350 | 0.525 | 0.745 |
| Layers 10, 11, 12 No Features | **0.436** | **0.244** | **0.480** | **0.370** | **0.554** | **0.758** |
| Layers 9, 10, 11, 12 No Features | 0.444 | 0.258 | 0.485 | 0.361 | 0.542 | 0.744 |
| Layers 4, 6, 8, 10, 12 No Features | 0.459 | 0.261 | 0.504 | 0.359 | 0.527 | 0.743 |
| Random Forest No Features | 0.484 | 0.294 | 0.527 | 0.385 | 0.508 | 0.710 |
| Gaussian Process No Features | 0.510 | 0.308 | 0.544 | 0.388 | 0.482 | 0.691 |
| Gradient Boosting No features | 0.527 | 0.358 | 0.570 | 0.427 | 0.463 | 0.642 |

Metrics for scaled values. The bold indicate the best mean results in terms of the metrics used with 5 fold validation.
Layer-wise fine-tuning, The attention weights are kept frozen for the unselected layers. 150,000 Compositions Pre-trained Model
No Features indicate that only the composition elements and their fractions are used as inputs to the model during training.

The consistent superiority of the pre-trained Transformer models across all evaluated scenarios can be attributed to several key factors. Firstly, the Transformer's architecture, equipped with self-attention mechanisms, excels at capturing complex relationships within the data. The self-attention layers allow the model to weigh the importance of each element in the composition relative to others, effectively modelling the interactions that govern material properties. Pre-training on a large and diverse dataset enhances this capability by exposing the model to a wide range of compositions and associated behaviours, enabling it to learn generalizable patterns that can be fine-tuned for specific tasks.

The benefits of transfer learning are evident in the performance gains observed with the pre-trained Transformer. By leveraging knowledge acquired during pre-training, the model requires less data and fewer epochs to converge during fine-tuning, improving generalization and reducing the risk of overfitting. This is particularly advantageous when fine-tuning datasets are limited in size or diversity, as the model can draw upon the rich representations learned during pre-training to make accurate predictions. The strategic fine-tuning of Transformer layers plays a significant role in optimizing performance for different material properties. The fact that fine-tuning deeper layers enhances UTS prediction suggests that UTS is more sensitive to high-level features and long-range dependencies within the material composition. Conversely, the necessity of fine-tuning all layers for optimal elongation prediction indicates that both local interactions and global patterns contribute to elongation, necessitating adjustments across the entire network.

The Transformer's robustness to limited feature sets underscores its strength in intrinsic feature extraction. Its ability to perform well without additional engineered features simplifies the modelling process and reduces the dependency on domain-specific expertise for feature engineering. This is particularly valuable in materials science, where complex interactions and high-dimensional data can make feature engineering challenging.

The combined plots in Figures 6 and 7 visualize the aggregated predictions and actual values from all validation sets across the K-Fold cross-validation process. In each fold, the model is trained on part of the data and evaluated on a separate validation (test) set, ensuring predictions are made on unseen data. By collecting these predictions and actual values from each fold, the combined plots provide a comprehensive view of the model's overall performance on the entire dataset as test data. This approach allows for a holistic assessment of the model's generalization ability and helps identify patterns or systematic errors across the full dataset.

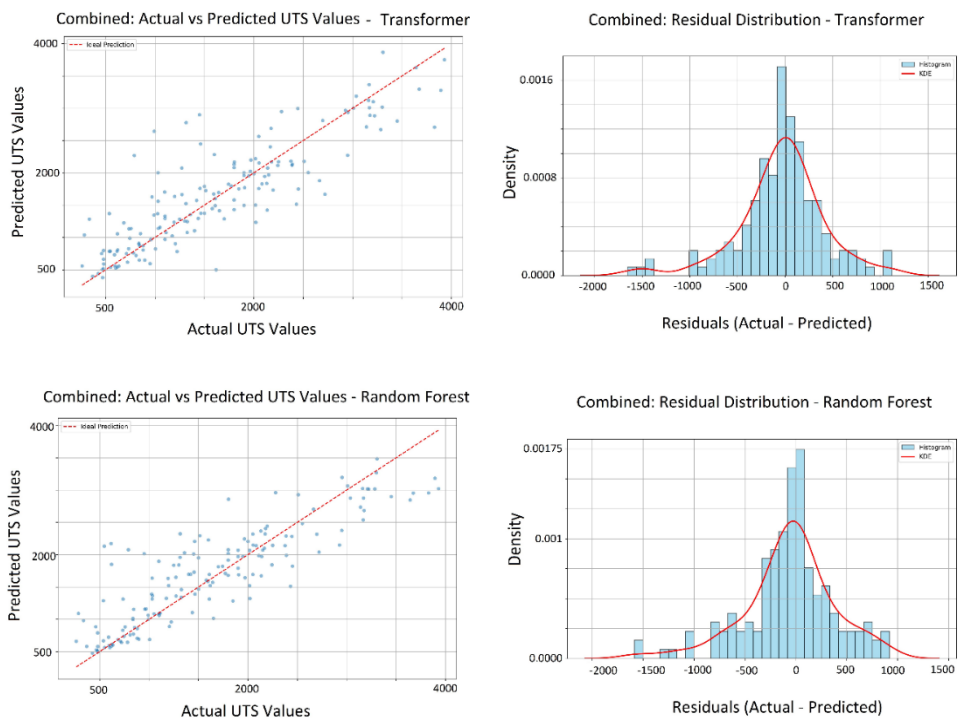

Figure 6. Scatter plot showing the relationship between actual and predicted values as well as the histogram of residuals (actual minus predicted values)

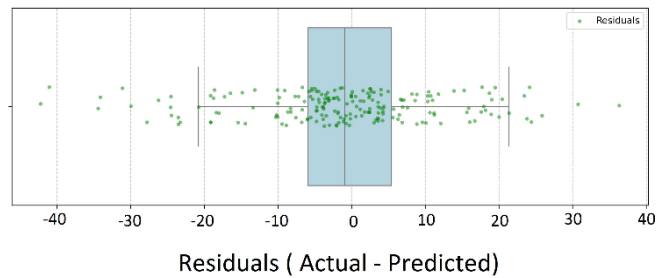

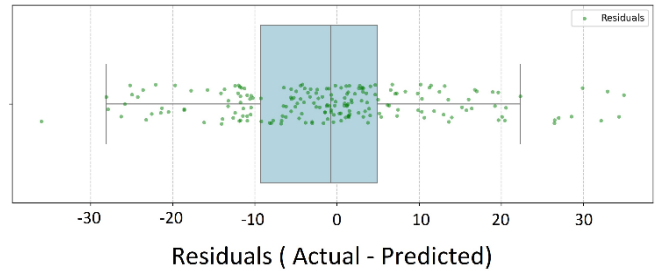

Figure 7. Box plot of residuals (actual minus predicted values)

An analysis of the "Actual vs. Predicted" scatter plots (Figure 6) for ultimate tensile strength (UTS) reveals that while both the Transformer and Random Forest models capture the general trend along the 45-degree line, the Transformer model displays a tighter clustering of data points around this line. This indicates a higher alignment between predicted and actual values, reflecting better accuracy and better generalization to unseen data. In contrast, the Random Forest model shows a broader dispersion of points around the line. The residual histograms reinforce these findings. The Transformer's residuals are sharply cantered around zero with a narrow distribution, indicating lower prediction variability and errors. Conversely, the Random Forest's residuals are more widely spread and less centered, pointing to greater variability and a higher average prediction error. Examining the residual box plots in Figure 7 provides further insight. The Transformer model's residuals exhibit a narrower interquartile range (IQR) and fewer outliers, signifying consistent low-error predictions across different validation folds. The Random Forest model displays a wider IQR and more outliers, indicating more significant deviations from actual values and less consistent performance. In summary, the Transformer model shows tighter clustering in the scatter plots, sharper and more cantered residual distribution, and narrower residual range.

**Model Interpretability**

The attention mechanisms of a pre-trained Transformer model can be analysed when processing a chemical composition input, specifically focusing on how the model attends to different elements within the material. It begins by tokenizing the input text of elemental compositions and mapping these tokens back to their corresponding chemical elements using offset mappings. The model's attention weights from the last layer are extracted and averaged across all heads to form a simplified attention matrix. This matrix represents how much the model's tokens attend to each other, essentially capturing the relationships between different elements in the composition. By grouping tokens corresponding to the same element and computing the average attention between these groups, a reduced attention matrix is created (Figures 8). This matrix is then symmetrized (averaged with its transpose) and visualized using a heatmap, with the diagonal masked to exclude self-attention. The resulting visualization highlights the strength of associations the model has learned between different elements, offering insights into potential chemical interactions and the model's focus when making predictions about properties like ultimate tensile strength (UTS).

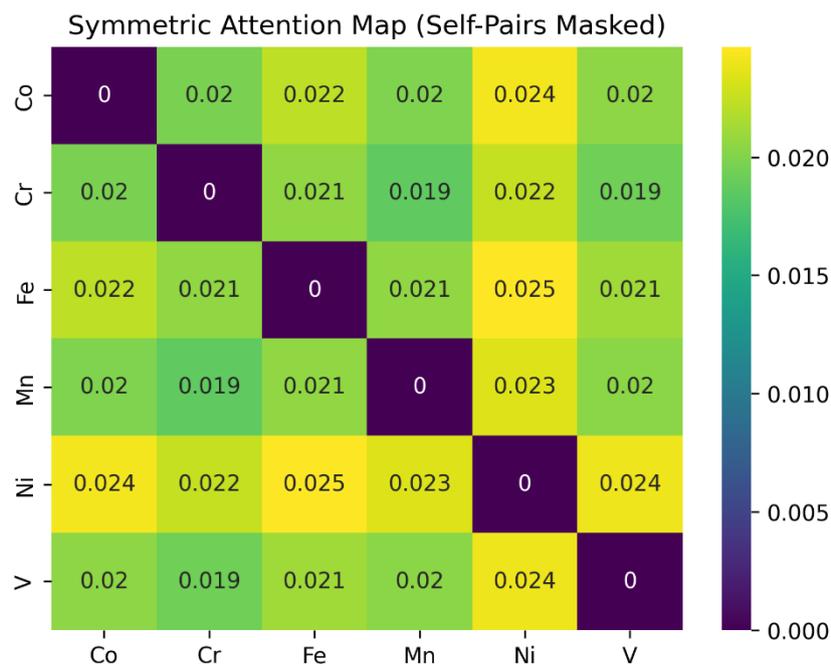

Figure 8 UTS attention maps for an unseen composition during training

Notably, the heatmap reveals high attention weights between Ni (Nickel) and Fe (Iron), as well as between Ni and Co (Cobalt), suggesting that the model has recognized these elements as having substantial interactions or associations within the alloy system. These observations align with established metallurgical principles regarding HEAs. Elements such as Ni, Fe, and Co are known to exhibit significant mutual solubility and tend to form stable face-centered cubic (FCC) solid solutions due to their similar atomic sizes and crystal structures. Their interactions contribute to the unique mechanical properties of HEAs, including enhanced ductility, toughness, and strength. The Transformer's high attention weights between these elements indicate that the model effectively captures these critical relationships, which are essential for accurate predictions of properties like ultimate tensile strength (UTS).

Furthermore, the heatmap shows moderate attention weights between Cr (Chromium) and Mn (Manganese), as well as between Cr and Fe. Chromium plays a crucial role in improving oxidation resistance and contributing to phase stability, while Manganese influences stacking fault energy and stabilizes certain phases, affecting deformation mechanisms such as twinning and slip. The model's attention to these element pairs suggests it recognizes their influence on the alloy's overall mechanical behaviour. By highlighting these interactions, the Transformer model demonstrates an ability to internalize complex metallurgical relationships that govern the properties of HEAs.

In summary, the model's focus on pairs like Ni–Fe and Ni–Co corresponds with known metallurgical phenomena, indicating that it captures the underlying physics and chemistry governing the alloy's behaviour. These findings are also supported by the literature where Ni content has a significant influence on hardness and crystal structure [45]. This alignment not only validates the model's predictive capabilities but also enhances its interpretability, offering valuable insights into how specific elemental interactions contribute to material properties. Such insights are instrumental in advancing materials informatics, as they bridge the gap between data-driven models and domain-specific knowledge, ultimately aiding in the design and discovery of new alloys with tailored properties.

**Conclusions**

The proposed pre-trained Transformer models demonstrate high performance in predicting material properties such as elongation and UTS, outperforming traditional machine learning models across various metrics. The advantages of large pre-training datasets, and strategic fine-tuning highlight the Transformer's capacity to capture complex material behaviours. These findings have significant implications for the field of materials informatics, suggesting that leveraging pre-trained Transformers can lead to more accurate and efficient predictions, ultimately accelerating the discovery and design of new materials. The ability to extract meaningful patterns from elemental compositions without reliance on engineered features also opens avenues for applying these models to other domains where data complexity and feature extraction pose significant challenges.

**Data Availability**

All data used in this work are publicly available. Original datasets could be found in https://github.com/CitrineInformatics/MPEA_dataset  Besides, the original and processed datasets used in this work are also available at https://github.com/SPS-Coatings/Language-Model-for-HEA

**Code Availability**

The codes developed for this work are available at https://github.com/SPS-Coatings/Language-Model-for-HEA

**References**

[1]     B. Cantor, I.T.H. Chang, P. Knight, A.J.B. Vincent, Microstructural development in equiatomic multicomponent alloys, Mater. Sci. Eng. A. 375 (2004) 213–218. https://doi.org/10.1016/j.msea.2003.10.257.

[2]     J.W. Yeh, S.K. Chen, S.J. Lin, J.Y. Gan, T.S. Chin, T.T. Shun, C.H. Tsau, S.Y. Chang, Nanostructured high-entropy alloys with multiple principal elements: Novel alloy design concepts and outcomes, Adv. Eng. Mater. 6 (2004) 299–303. https://doi.org/10.1002/adem.200300567.

[3]     D.B. Miracle, O.N. Senkov, A critical review of high entropy alloys and related concepts, Acta Mater. 122 (2017) 448–511. https://doi.org/10.1016/j.actamat.2016.08.081.

[4]     O.F. Dippo, K.S. Vecchio, A universal configurational entropy metric for high-entropy materials, Scr. Mater. (2021). https://doi.org/10.1016/j.scriptamat.2021.113974.

[5]     S. Marik, K. Motla, M. Varghese, K.P. Sajilesh, D. Singh, Y. Breard, P. Boullay, R.P. Singh, Superconductivity in a new hexagonal high-entropy alloy, Phys. Rev. Mater. (2019). https://doi.org/10.1103/PhysRevMaterials.3.060602.

[6]     J.W. Yeh, Recent progress in high-entropy alloys, Ann. Chim. Sci. Des Mater. 31 (2006) 633–648. https://doi.org/10.3166/acsm.31.633-648.

[7]  B. Wang, C. Yang, D. Shu, B. Sun, A Review of Irradiation-Tolerant Refractory High-Entropy Alloys, Metals (Basel). (2024). https://doi.org/10.3390/met14010045.

[8]  C. Han, Q. Fang, Y. Shi, S.B. Tor, C.K. Chua, K. Zhou, Recent Advances on High-Entropy Alloys for 3D Printing, Adv. Mater. (2020). https://doi.org/10.1002/adma.201903855.

[9]  X. Chang, M. Zeng, K. Liu, L. Fu, Phase Engineering of High-Entropy Alloys, Adv. Mater. (2020). https://doi.org/10.1002/adma.201907226.

[10]  Y. Zhang, T.T. Zuo, Z. Tang, M.C. Gao, K.A. Dahmen, P.K. Liaw, Z.P. Lu, Microstructures and properties of high-entropy alloys, Prog. Mater. Sci. (2014). https://doi.org/10.1016/j.pmatsci.2013.10.001.

[11]  J. Gao, J. Zhong, G. Liu, S. Yang, B. Song, L. Zhang, Z. Liu, A machine learning accelerated distributed task management system (Malac-Distmas) and its application in high-throughput CALPHAD computation aiming at efficient alloy design, Adv. Powder Mater. (2022). https://doi.org/10.1016/j.apmate.2021.09.005.

[12]  R. Feng, P.K. Liaw, M.C. Gao, M. Widom, First-principles prediction of high-entropy-alloy stability, Npj Comput. Mater. 3 (2017). https://doi.org/10.1038/s41524-017-0049-4.

[13]  X. Yang, Z. Wang, X. Zhao, J. Song, M. Zhang, H. Liu, MatCloud: A high-throughput computational infrastructure for integrated management of materials simulation, data and resources, Comput. Mater. Sci. (2018). https://doi.org/10.1016/j.commatsci.2018.01.039.

[14]  R. Li, L. Xie, W.Y. Wang, P.K. Liaw, Y. Zhang, High-Throughput Calculations for High-Entropy Alloys: A Brief Review, Front. Mater. (2020). https://doi.org/10.3389/fmats.2020.00290.

[15]  C. Zhang, X. Jiang, R. Zhang, X. Wang, H. Yin, X. Qu, Z.K. Liu, High-throughput thermodynamic calculations of phase equilibria in solidified 6016 Al-alloys, Comput. Mater. Sci. (2019). https://doi.org/10.1016/j.commatsci.2019.05.022.

[16]  K. Song, J. Xing, Q. Dong, P. Liu, B. Tian, X. Cao, Optimization of the processing parameters during internal oxidation of Cu-Al alloy powders using an artificial neural network, Mater. Des. (2005). https://doi.org/10.1016/j.matdes.2004.06.002.

[17]  Y. Sun, W.D. Zeng, Y.Q. Zhao, Y.L. Qi, X. Ma, Y.F. Han, Development of constitutive relationship model of Ti600 alloy using artificial neural network, Comput. Mater. Sci. (2010). https://doi.org/10.1016/j.commatsci.2010.03.007.

[18]  J. Su, Q. Dong, P. Liu, H. Li, B. Kang, Prediction of properties in thermomechanically treated Cu-Cr-Zr alloy by an artificial neural network, J. Mater. Sci. Technol. (2003).

[19]  S. Malinov, W. Sha, J.J. McKeown, Modelling the correlation between processing parameters and properties in titanium alloys using artificial neural network, Comput. Mater. Sci. (2001). https://doi.org/10.1016/S0927-0256(01)00160-4.

[20]  J. Warde, D.M. Knowles, Use of neural networks for alloy design, ISIJ Int. (1999). https://doi.org/10.2355/isijinternational.39.1015.

[21] Y. Sun, W.D. Zeng, Y.Q. Zhao, X.M. Zhang, Y. Shu, Y.G. Zhou, Modeling constitutive relationship of Ti40 alloy using artificial neural network, Mater. Des. (2011). https://doi.org/10.1016/j.matdes.2010.10.004.

[22] S.K. Dewangan, S. Samal, V. Kumar, Microstructure exploration and an artificial neural network approach for hardness prediction in AlCrFeMnNiWx High-Entropy Alloys, J. Alloys Compd. (2020). https://doi.org/10.1016/j.jallcom.2020.153766.

[23] J. Wang, H. Kwon, H.S. Kim, B.J. Lee, A neural network model for high entropy alloy design, Npj Comput. Mater. (2023). https://doi.org/10.1038/s41524-023-01010-x.

[24] W.C. Lu, X.B. Ji, M.J. Li, L. Liu, B.H. Yue, L.M. Zhang, Using support vector machine for materials design, Adv. Manuf. (2013). https://doi.org/10.1007/s40436-013-0025-2.

[25] W. Zhang, P. Li, L. Wang, F. Wan, J. Wu, L. Yong, Explaining of prediction accuracy on phase selection of amorphous alloys and high entropy alloys using support vector machines in machine learning, Mater. Today Commun. (2023). https://doi.org/10.1016/j.mtcomm.2023.105694.

[26] N.H. Chau, M. Kubo, L.V. Hai, T. Yamamoto, Support Vector Machine-Based Phase Prediction of Multi-Principal Element Alloys, Vietnam J. Comput. Sci. (2023). https://doi.org/10.1142/S2196888822500312.

[27] F. Tancret, I. Toda-Caraballo, E. Menou, P.E.J. Rivera Díaz-Del-Castillo, Designing high entropy alloys employing thermodynamics and Gaussian process statistical analysis, Mater. Des. (2017). https://doi.org/10.1016/j.matdes.2016.11.049.

[28] S.M. Park, T. Lee, J.H. Lee, J.S. Kang, M.S. Kwon, Gaussian process regression-based Bayesian optimization of the insulation-coating process for Fe-Si alloy sheets, J. Mater. Res. Technol. (2023). https://doi.org/10.1016/j.jmrt.2022.12.171.

[29] D. Khatamsaz, B. Vela, R. Arróyave, Multi-objective Bayesian alloy design using multi-task Gaussian processes, Mater. Lett. (2023). https://doi.org/10.1016/j.matlet.2023.135067.

[30] F. Tancret, Computational thermodynamics, Gaussian processes and genetic algorithms: Combined tools to design new alloys, Model. Simul. Mater. Sci. Eng. (2013). https://doi.org/10.1088/0965-0393/21/4/045013.

[31] T.J. Sabin, C.A.L. Bailer-Jones, P.J. Withers, Accelerated learning using Gaussian process models to predict static recrystallization in an Al-Mg alloy, Model. Simul. Mater. Sci. Eng. (2000). https://doi.org/10.1088/0965-0393/8/5/304.

[32] H. Liu, Y.S. Ong, X. Shen, J. Cai, When Gaussian Process Meets Big Data: A Review of Scalable GPs, IEEE Trans. Neural Networks Learn. Syst. (2020). https://doi.org/10.1109/TNNLS.2019.2957109.

[33] R. Ghouchan Nezhad Noor Nia, M. Jalali, M. Houshmand, A Graph-Based k-Nearest Neighbor (KNN) Approach for Predicting Phases in High-Entropy Alloys, Appl. Sci. (2022). https://doi.org/10.3390/app12168021.

[34] Ö.F. Ertuğrul, M.E. Tağluk, A novel version of k nearest neighbor: Dependent nearest neighbor, Appl. Soft Comput. J. (2017). https://doi.org/10.1016/j.asoc.2017.02.020.

[35] J. Zhang, J. Wu, A. Yin, Z. Xu, Z. Zhang, H. Yu, Y. Lu, W. Liao, L. Zheng, Grain size characterization of Ti-6Al-4V titanium alloy based on laser ultrasonic random forest regression, Appl. Opt. (2023). https://doi.org/10.1364/ao.479323.

[36] Z. Zhang, Z. Yang, W. Ren, G. Wen, Random forest-based real-time defect detection of Al alloy in robotic arc welding using optical spectrum, J. Manuf. Process. (2019). https://doi.org/10.1016/j.jmapro.2019.04.023.

[37] Simple Introduction to Convolutional Neural Networks | by Matthew Stewart, PhD Researcher | Towards Data Science, (n.d.). https://towardsdatascience.com/simple-introduction-to-convolutional-neural-networks-cdf8d3077bac (accessed September 6, 2021).

[38] X. Wang, N.D. Tran, S. Zeng, C. Hou, Y. Chen, J. Ni, Element-wise representations with ECNet for material property prediction and applications in high-entropy alloys, Npj Comput. Mater. (2022). https://doi.org/10.1038/s41524-022-00945-x.

[39] S. Feng, H. Zhou, H. Dong, Application of deep transfer learning to predicting crystal structures of inorganic substances, Comput. Mater. Sci. (2021). https://doi.org/10.1016/j.commatsci.2021.110476.

[40] A. Vaswani, N. Shazeer, N. Parmar, J. Uszkoreit, L. Jones, A.N. Gomez, Ł. Kaiser, I. Polosukhin, Attention is all you need, in: Adv. Neural Inf. Process. Syst., 2017. https://doi.org/10.48550/arXiv.1706.03762%0A.

[41] A. Radford, K. Narasimhan, T. Salimans, I. Sutskever, Improving Language Understanding by Generative Pre-Training, OpenAI.Com. (2018) 1–12. https://cdn.openai.com/research-covers/language-unsupervised/language_understanding_paper.pdf.

[42] C. Xu, Y. Wang, A. Barati Farimani, TransPolymer: a Transformer-based language model for polymer property predictions, Npj Comput. Mater. (2023). https://doi.org/10.1038/s41524-023-01016-5.

[43] S. Kamnis, Introducing pre-trained transformers for high entropy alloy informatics, Mater. Lett. (2024). https://doi.org/10.1016/j.matlet.2024.135871.

[44] J. Devlin, M.W. Chang, K. Lee, K. Toutanova, BERT: Pre-training of deep bidirectional transformers for language understanding, in: NAACL HLT 2019 - 2019 Conf. North Am. Chapter Assoc. Comput. Linguist. Hum. Lang. Technol. - Proc. Conf., 2019.

[45] S. González, A.K. Sfikas, S. Kamnis, C.G. Garay-Reyes, A. Hurtado-Macias, R. Martínez-Sánchez, Wear resistant CoCrFeMnNi0.8V high entropy alloy with multi length-scale hierarchical microstructure, Mater. Lett. (2023). https://doi.org/10.1016/j.matlet.2022.133504.

# Appendix A

**General Symbols**

- $S$ : A chemical composition string (e.g., "Co1.2 Fe0.8 Ni1").
- $\mathcal{A}$ : The set of all chemical element symbols.
- $N$ : Number of samples in the dataset or number of Transformer layers, depending on context.
- $L$ : Length of the input token sequence.
- $h$ : Number of attention heads in the Transformer model.
- $d$: Dimensionality of the embedding space.
- $d_k$ : Dimensionality of the query and key vectors, typically $d_k = d/h$.
- $d_{ff}$ : Dimensionality of the feed-forward network's inner layer.
- $\sigma$ : Activation function (e.g., GELU, ReLU) or standard deviation, depending on context.
- $\mu$ : Mean value, often of a feature or target variable.
- $\eta$ : Learning rate used in optimization.
- $\lambda$ : Weight decay coefficient for regularization.
- $\theta$ : Model parameters.
- $\gamma$ : Gradient clipping threshold.

**Data Preprocessing Symbols**

- $E_i$ : The $i$-th chemical element symbol extracted from a composition.
- $f_i$ : Fractional abundance (stoichiometric coefficient) of element $E_i$.
- $\mathcal{E}(S)$ : Set of element-fraction pairs extracted from composition $S$ :
$$\mathcal{E}(S) = \{(E_i, f_i) \mid i = 1, \dots, N_S\}$$
- $N_S$ : Number of unique elements in composition $S$.
- $\mathcal{E}'(S)$ : Alphabetically sorted set of element-fraction pairs from $S$.
- $T_i$ : Token formed by concatenating $E_i$ and $f_i$ :
$$T_i = E_i \circ f_i$$
- $\mathcal{T}(S)$ : Sequence of tokens for composition $S$ :
$$\mathcal{T}(S) = [T_1, T_2, \dots, T_{N_S}]$$
- $x_j$ : The $j$-th numerical feature (e.g., temperature, pressure).
- $\tilde{x}_j$: Normalized numerical feature:
$$\tilde{x}_j = \frac{x_j - \mu_{x_j}}{\sigma_{x_j}}$$
- $C_S$ : Combined features for sample $S$, including tokenized elements and numerical features:
$$C_S = \mathcal{T}(S) \cup \{\tilde{x}_1, \tilde{x}_2, \dots, \tilde{x}_M\}$$

- $y$ : Original target variable (e.g., hardness, ultimate tensile strength).
- $\tilde{y}$ : Normalized target variable:

$$\tilde{y} = \frac{y - \mu_{\text{train}}}{\sigma_{\text{train}}}$$

- $\mu_{\text{train}}, \sigma_{\text{train}}$ : Mean and standard deviation of the target variable in the training set.

**Tokenization and Embedding Symbols**

- $\mathbf{t} = [t_1, t_2, \ldots, t_L]$ : Tokenized input sequence after applying BERT tokenizer.
- V: Vocabulary of the tokenizer.
- $\text{ID}(t_i)$ : Vocabulary index of token $t_i$.
- $\mathbf{t}_{\text{ids}} = [\text{ID}(t_1), \text{ID}(t_2), \ldots, \text{ID}(t_L)]$ : Sequence of token IDs.
- E: Token embedding matrix of size $|V| \times d$.
- $\mathbf{1}_{t_i}$ : One-hot encoding vector for token $t_i$.
- $\mathbf{e}_{t_i}$ : Token embedding for $t_i$ :

$$\mathbf{e}_{t_i} = \mathbf{E} \cdot \mathbf{1}_{t_i}$$

- $\mathbf{E}_t = [\mathbf{e}_{t_1}, \mathbf{e}_{t_2}, \ldots, \mathbf{e}_{t_L}]$ : Sequence of token embeddings.
- $\mathbf{e}_{\text{pos}_i}$ : Positional embedding for position $i$.
- $\mathbf{E}_{\text{pos}} = [\mathbf{e}_{\text{pos}_1}, \mathbf{e}_{\text{pos}_2}, \ldots, \mathbf{e}_{\text{pos}_L}]$ : Sequence of positional embeddings.
- $\mathbf{e}_{\text{seg}}$ : Segment embedding (constant vector for all positions in single-sequence tasks).
- $\mathbf{E}_{\text{seg}} = [\mathbf{e}_{\text{seg}}, \mathbf{e}_{\text{seg}}, \ldots, \mathbf{e}_{\text{seg}}]$ : Sequence of segment embeddings.
- $\mathbf{H}^{(0)}$ : Input embeddings to the Transformer encoder:

$$\mathbf{H}^{(0)} = \mathbf{E}_t + \mathbf{E}_{\text{pos}} + \mathbf{E}_{\text{seg}}$$

- $\mathbf{H}^{(l)}$ : Hidden states at layer $l$ of the Transformer encoder.

**Transformer Encoder Symbols**
**Multi-Head Self-Attention**

- $\mathbf{Q}_i, \mathbf{K}_i, \mathbf{V}_i$ : Query, key, and value vectors at position $i$ :

$$\mathbf{Q}_i = \mathbf{W}_Q \mathbf{H}_i^{(l-1)}$$
$$\mathbf{K}_i = \mathbf{W}_K \mathbf{H}_i^{(l-1)}$$
$$\mathbf{V}_i = \mathbf{W}_V \mathbf{H}_i^{(l-1)}$$

- $\mathbf{W}_Q, \mathbf{W}_K, \mathbf{W}_V \in \mathbb{R}^{d \times d_k}$ : Projection matrices for queries, keys, and values.
- $\mathbf{Q}_i^{(h)}, \mathbf{K}_i^{(h)}, \mathbf{V}_i^{(h)}$ : Query, key, and value vectors for head $h$.
- $\alpha_{ij}^{(h)}$: Attention weight from position $i$ to position $j$ in head $h$ :

$$\alpha_{ij}^{(h)} = \text{softmax}\left(\frac{\mathbf{Q}_i^{(h)} \cdot \left(\mathbf{K}_j^{(h)}\right)^\top}{\sqrt{d_k}}\right)$$

- $\mathbf{Z}_i^{(h)}$ : Output of attention head $h$ at position $i$ :

$$\mathbf{Z}_i^{(h)} = \sum_{j=1}^{L} \alpha_{ij}^{(h)} \mathbf{V}_j^{(h)}$$

- $\mathbf{Z}_i$ : Concatenated output of all attention heads at position $i$ :

$$\mathbf{Z}_i = \mathbf{W}_O \left[\mathbf{Z}_i^{(1)} \,\|\, \mathbf{Z}_i^{(2)} \,\|\, \ldots \,\|\, \mathbf{Z}_i^{(h)}\right]$$

- $\mathbf{W}_O \in \mathbb{R}^{d \times d}$ : Output projection matrix after concatenation.
- $\|$: Concatenation operator for vectors.
- $\tilde{\mathbf{H}}_i^{(l)}$ : Output after residual connection and layer normalization:

$$\tilde{\mathbf{H}}_i^{(l)} = \text{LayerNorm}\left(\mathbf{H}_i^{(l-1)} + \mathbf{Z}_i\right)$$

**Position-Wise Feed-Forward Network**

- $\mathbf{F}_i^{(l)}$ : Output of the feed-forward network at layer $l$, position $i$ :

$$\mathbf{F}_i^{(l)} = \sigma\left(\mathbf{W}_1 \tilde{\mathbf{H}}_i^{(l)} + \mathbf{b}_1\right) \mathbf{W}_2 + \mathbf{b}_2$$

- $\mathbf{W}_1 \in \mathbb{R}^{d \times d_d}, \mathbf{W}_2 \in \mathbb{R}^{d_{II} \times d}$ : Weights of the feed-forward network.
- $\mathbf{b}_1 \in \mathbb{R}^{d_{\tilde{\pi}}}, \mathbf{b}_2 \in \mathbb{R}^{d}$ : Biases of the feed-forward network.
- $\sigma$ : Activation function (e.g., GELU, ReLU).
- $\mathbf{H}_i^{(l)}$ : Output after the second residual connection and layer normalization:

$$\mathbf{H}_i^{(l)} = \text{LayerNorm}\left(\tilde{\mathbf{H}}_i^{(l)} + \mathbf{F}_i^{(l)}\right)$$

**Regression Head Symbols**

- $\mathbf{h}_{\text{CLS}} = \mathbf{H}_1^{(N)}$ : Final hidden state of the [CLS] token after $N$ Transformer layers.
- $\mathbf{W}_{\text{reg}} \in \mathbb{R}^d$ : Weights of the regression head.
- $b_{\text{reg}} \in \mathbb{R}$ : Bias term of the regression head.
- $\hat{y}$ : Predicted normalized target value:

$$\hat{y} = \mathbf{W}_{\text{reg}}^\top \mathbf{h}_{\text{CLS}} + b_{\text{reg}}$$

- $\hat{y}_i^{(\text{orig})}$ : Predicted target value in original scale for sample $i$ :

$$\hat{y}_i^{(\text{orig})} = \hat{y}_i \sigma_{\text{train}} + \mu_{\text{train}}$$

**Loss Function and Evaluation Metrics**

- $\mathcal{L}(\theta)$ : Mean Squared Error (MSE) loss function:

$$\mathcal{L}(\theta) = \frac{1}{N} \sum_{i=1}^{N} (\hat{y}_i - \tilde{y}_i)^2$$

- MSE: Mean Squared Error metric:

$$\text{MSE} = \frac{1}{N_{\text{val}}} \sum_{i=1}^{N_{\text{ral}}} (\hat{y}_i - \tilde{y}_i)^2$$

- MAE: Mean Absolute Error metric:

$$\text{MAE} = \frac{1}{N_{\text{val}}} \sum_{i=1}^{N_{\text{val}}} |\hat{y}_i - \tilde{y}_i|$$

- $R^2$ : Coefficient of determination:

$$R^2 = 1 - \frac{\sum_{i=1}^{N_{\text{vel}}} (\hat{y}_i - \tilde{y}_i)^2}{\sum_{i=1}^{N_{\text{vel}}} (\tilde{y}_i - \bar{y}_{\text{val}})^2}$$

- $\bar{y}_{\text{val}}$ : Mean of the true normalized target values in the validation set.

**Residuals and Analysis Symbols**

- $r_i$ : Residual for sample $i$ :

$$r_i = y_i - \hat{y}_i^{(\text{orig})}$$

- M: Attention mask used to indicate valid tokens versus padding tokens.
- LayerNorm( **x** ): Layer normalization function applied to vector **x** :

$$\text{LayerNorm}(\mathbf{x}) = \frac{\mathbf{x} - \mu_{\mathbf{x}}}{\sigma_{\mathbf{x}}} \cdot \gamma + \beta$$

- $\mu_{\mathbf{x}}, \sigma_{\mathbf{x}}$ : Mean and standard deviation of components of **x**.
- $\gamma, \beta$ : Learnable scaling and shifting parameters.

**Optimization Symbols**

- $\nabla_\theta \mathcal{L}(\theta)$ : Gradient of the loss function with respect to model parameters.
- $\theta_t$ : Model parameters at iteration $t$.
- $\theta_{t+1}$ : Updated model parameters after applying optimization step.
- $\lambda\theta$ : Weight decay term for regularization.
- $\frac{\partial \hat{\theta}_1}{\partial \theta}$ : Gradient of the prediction with respect to model parameters.

**Miscellaneous Symbols**

- $\text{clip}(\theta_t, -\gamma, \gamma)$ : Gradient clipping operation to keep parameters within $[-\gamma, \gamma]$.

- softmax($\cdot$) : Softmax function applied to attention scores to obtain attention weights.
- FFN ($\cdot$) : Position-wise feed-forward network function.
- $\mathbf{H}_i^{(l-1)}$ : Hidden state at layer $l-1$, position $i$.
- $\mathbf{H}^{(N)}$ : Final hidden states after $N$ Transformer layers.

Note: All vectors are assumed to be column vectors unless specified otherwise. Matrices and vectors are represented in boldface, while scalars are in regular typeface. The operators and functions are standard in linear algebra and machine learning contexts.